**Anisotropic Nodal-Line-Derived Large Anomalous Hall Conductivity in ZrMnP and HfMnP**

*Sukriti Singh, Jonathan Noky, Shaileyee Bhattacharya, Praveen Vir, Yan Sun, Nitesh Kumar, Claudia Felser, and Chandra Shekhar\**

S. Singh, Dr. J. Noky, Dr. Y. Sun, Prof. C. Felser, Dr. C. Shekhar
Max Planck Institute for Chemical Physics of Solids, 01187 Dresden, Germany
E-mail: shekhar@cpfs.mpg.de
S. Bhattacharya
Max Planck Institute for Chemical Physics of Solids, 01187 Dresden, Germany
Present address: Paul Scherrer Institut, CH- 5232 Villigen-PSI, Switzerland
Dr. P. Vir
Max Planck Institute for Chemical Physics of Solids, 01187 Dresden, Germany
Present address: Diffraction group, Institut Laue-Langevin, 71 Avenue des Martyrs, 38000, Grenoble, France
Dr. N. Kumar
Max Planck Institute for Chemical Physics of Solids, 01187 Dresden, Germany
Present address: S. N. Bose National Centre for Basic Sciences, Salt Lake City, Kolkata - 700 106, India

The nontrivial band structure of semimetals has attracted substantial research attention in condensed matter physics and materials science in recent years owing to its intriguing physical properties. Within this class, a group of non-trivial materials known as nodal-line semimetals is particularly important. Nodal-line semimetals exhibit the potential effects of electronic correlation in nonmagnetic materials, whereas they enhance the contribution of the Berry curvature in magnetic materials, resulting in high anomalous Hall conductivity (AHC). In this study, two ferromagnetic compounds, namely ZrMnP and HfMnP, are selected, wherein the abundance of mirror planes in the crystal structure ensures gapped nodal lines at the Fermi energy. These nodal lines result in one of the largest AHC values of 2840 $\Omega^{-1}$cm$^{-1}$, with a high anomalous Hall angle of 13.6% in these compounds. First-principles calculations provide a clear and detailed understanding of nodal line-enhanced AHC. Our finding suggests a guideline for searching large AHC compounds.

# 1. Introduction

Topological materials have attracted substantial attention owing to their nontrivial electronic band structures that offer the potential for revolutionary device applications. A nontrivial band topology appears when band inversion occurs in momentum space; that is, the conduction band is below the valence band with respect to their natural order. This inversion may occur in several manners, and the corresponding wave function of each band twists and induces a finite Berry phase that is associated with the Berry curvature (BC). In addition to the accidental touching of bands at a node, they may also form a line, and such compounds are known as nodal-line compounds. The materials may be both magnetic and nonmagnetic, in which the nodal lines are protected by mirror symmetry,[1-7] and consequently, a drum head-like topological surface state exists.[1-4] In ferromagnets, all bands are usually singly degenerate but in the presence of mirror planes, they may be doubly degenerate in the form of a Dirac nodal line. In such materials, the magnetism and topology are entangled, and depending on the applied magnetic field direction, the degeneracy of the nodal line is lifted and a nontrivial gap is opened.[8, 9] The BC sum over such gapped lines is enhanced significantly. The BC that is associated with nontrivial bands as a source of anomalous Hall conductivity (AHC) has recently been recognized in various compounds; for example, the chiral antiferromagnets $Mn_3Sn$[10] and $Mn_3Ge$,[11] kagome lattice ferromagnet $Co_3Sn_2S_2$,[12] and nodal-line ferromagnets $Co_2MnZ$ (Z = Ga, Al)[8, 13, 14] and MnAlGe.[15] The presence of nodal lines has been revealed experimentally by spectroscopy.[15, 16] Among these, $Co_2MnZ$ has exhibited the record AHC value of 1600 to 2000 $\Omega^{-1}cm^{-1}$ at 2 K and many more compounds from the same family are awaiting experimental realization, which suggests the crucial roles of mirror planes.[9] Apart from the nontrivial band topology, an important factor for the occurrence of large AHC is spin-orbit coupling (SOC). We observed an AHC of 2000 $\Omega^{-1}cm^{-1}$ for ZrMnP and 2840 $\Omega^{-1}cm^{-1}$ for HfMnP. In this work, our approach was to study the AHC in a system with large SOC and that contained ample mirror planes in the crystal structures.

Transition metal pnictides are of significant interest as they possess both a high ferromagnetic transition temperature ($T_C$) above room temperature and large magnetic anisotropy. ZrMnP and HfMnP are particularly important compounds, which crystallize in a TiNiSi-type orthorhombic structure with space group (SG) *Pnma* (No. 62). Single-crystal x-ray analyses demonstrated the lattice parameters to be $a$ = 3.64 Å, $b$ = 6.45 Å, and $c$ = 7.53 Å for ZrMnP and $a$ = 3.61 Å, $b$ = 6.38 Å, and $c$ = 7.47 Å for HfMnP. Both ZrMnP and HfMnP are ferromagnets and their observed $T_C$ values were 320 and 370 K, respectively (**Figure 1(d)**), which match well with the previous report.[17] Only the Mn atoms contributed to the magnetism (as indicated in Figure 1(a)). The crystals of ZrMnP and HfMnP grew in a needle shape along the *b*-axis with well-defined facets, as demonstrated by the Laue x-ray diffraction (Figure 1(b)). However, the *a*- and *c*-axes were the easy and difficult directions for the magnetic moments, respectively, resulting in large magnetic anisotropy.[17] The measured values of the saturation magnetization at 2 K were 1.8μ$_B$/f.u. for ZrMnP and 2.0μ$_B$/f.u. for HfMnP, which is consistent with previously reported value.[17] Our first-principles calculations revealed that these compounds possess Dirac nodal lines, which are protected by the various mirror planes at different energies.

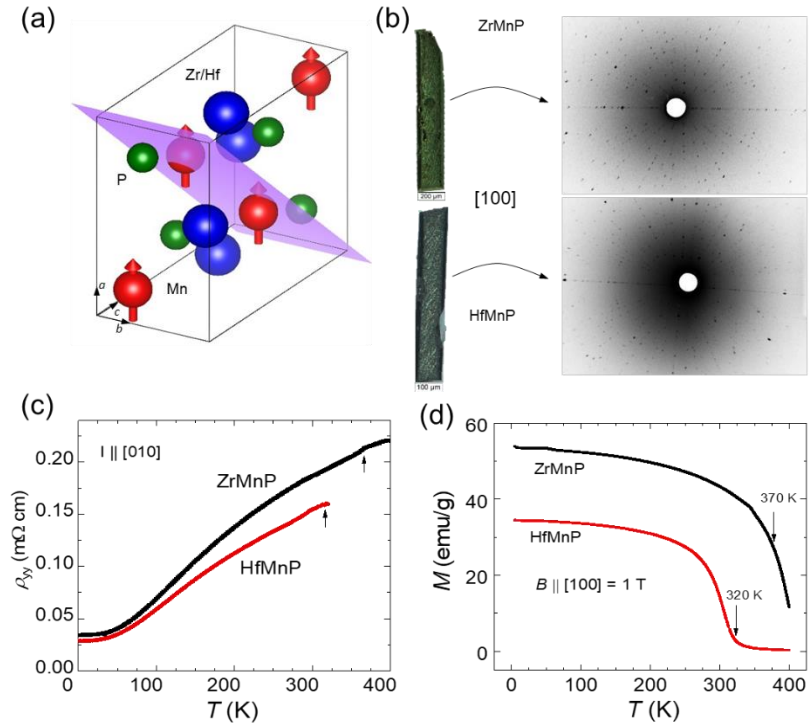

**Figure 1.** Crystal structure, Laue pattern, resistivity, and magnetic measurements of ZrMnP and HfMnP. (a) Orthorhombic unit cell of compounds with SG *Pnma*. The highlighted (101) is one of the mirror planes. (b) [100] facet of as-grown crystals and their corresponding Laue patterns, where the needle direction is [010]. (c) Temperature-dependent longitudinal resistivity $\rho_{yy}$ along [010], indicating a metallic character with the kinks (indicated by arrows) corresponding to the ferromagnetic transition temperature ($T_C$) of the compounds. (d) Temperature-dependent magnetization along [100] with $T_C$ indicated by arrows.

## 2. Results and Discussion

The zero-field longitudinal resistivity $\rho_{yy}$ (Figure 1(c)) of both compounds increased as the temperature increased, indicating metallic behavior. For the current $I \parallel [010]$, the value of $\rho_{yy}$ at 2 K was $3.4 \times 10^{-5}$ Ωcm for ZrMnP and $2.9 \times 10^{-5}$ Ωcm for HfMnP. Both compounds exhibited similar values of the residual resistivity ratio [RRR = $\rho_{yy}$ (300 K)/$\rho_{yy}$(2 K)], which was 5.5, reflecting high quality of the crystals. The kinks in their respective resistivity measurements corresponded to the magnetic transition. [17] A magnetic transition from the ferromagnetic to paramagnetic state was observed in the temperature-dependent magnetic measurements with a magnetic field $B = 1$ T along [100], as illustrated in Figure 1(b). The ZrMnP exhibited a relatively higher $T_C$ (370 K) than that of HfMnP (320 K), which was consistent with the resistivity measurements. Throughout this manuscript, the *a*-, *b*-, and *c*-axes of the compounds are equivalent to the *x*-, *y*-, and *z*-directions in the measurements, respectively.

As illustrated in Figure 1, the grown crystals of both compounds had a needle shape, which provided the opportunity to perform Hall measurements without any additional work of cutting the crystals in a Hall bar. The magnetic field-dependent Hall resistivity $\rho_{yz}$ was measured at various temperatures in the range of -9 to 9 T. The observed behavior of $\rho_{yz}$ is presented in **Figure 2(a)** and **(b)**, where the current was passed along the *b*-axis and the magnetic field was applied along the *a*-axis; that is, the easy axis of magnetization. The sign of the measured Hall resistivity data for the ZrMnP indicates that the holes constituted the majority of the charge carriers, whereas in the case of HfMnP, the majority of charge carriers were electrons. These findings are in good agreement with the numerical results from the density functional theory (DFT) calculations presented in **Figure 4(a)** and **(b)**. A hole pocket exists around the U-point in the first Brillouin zone, which cuts the Fermi energy ($E_F$) for ZrMnP but not for HfMnP. An anomaly is observed in the $\rho_{yz}$ measurements, which is attributed to the anomalous Hall effect (AHE) that normally appears in metallic ferromagnets; that is, a sharp increase at a lower field, followed by saturation with a further increase in the field. Moreover, it exhibits a resemblance with the magnetization curve; however, its saturation value decreases with a decrease in the temperature. The measured $\rho_{yx}$ in ferromagnets is usually defined as a combination of two terms: $\rho_{xy}(T) = \rho_{OHE}(T) + \rho_{AHE}(T)$, where $\rho_{OHE}$ is the contribution from the ordinary Hall effect that arises from the Lorentz force acting on the charge carriers, whereas $\rho_{AHE}$ is the anomalous Hall contribution, which is unique to magnetic samples.[18] The value of $\rho_{AHE}$ at 2 K is 2.5 µΩcm for ZrMnP and 2.3 µΩcm for HfMnP (Figure 2(a) and (d)), which indicates an increasing trend, and reaches 17.2 μΩcm and 16 μΩcm, respectively, at 250 K. The $\rho_{AHE}$ value at a fixed temperature was estimated by interpolating the high field value of $\rho_{yz}$ to the zero-field as the y-intercept. Similarly, the AHC $\sigma_{AHC}$ was estimated from the Hall conductivity ($\sigma_{zy} = \frac{\rho_{yz}}{\rho_{yz}^2 + \rho_{yy}^2}$) as the y-intercept. The $\sigma_{AHC}$ at 2 K is 2000 $\Omega^{-1}$cm$^{-1}$ for ZrMnP and 2840 $\Omega^{-1}$cm$^{-1}$ for HfMnP (Figure 2(b) and (e)), which is observable up to their transition temperature (Figure 2(c) and (f)). The total $\sigma_{AHC}$ has both intrinsic and extrinsic contributions from different mechanisms. The intrinsic contribution originates from the electronic band structure (the BC), whereas the extrinsic contribution arises from the skew scattering or side-jump effect. To define these contributions empirically, the temperature-dependent estimated $\rho_{xy}^A$ can be expressed as $\rho_{xy}^A(T) = \alpha \rho_{xx0} + \beta \rho_{xx0}^2 + \gamma \rho_{xx}^2(T)$, where $\rho_{xx0}$ is the residual longitudinal resistivity.[18] The first and second terms represent the extrinsic contributions from the skew scattering and side-jump, respectively, whereas the final term denotes the intrinsic contribution from the BC that is associated with nontrivial bands. To extract the intrinsic contribution, we plotted $\rho_{yz}^A$ as a function of $\rho_{yy}^2$ at different temperatures. The slope $\gamma$ of the straight line directly provides the estimation of the intrinsic value of AHC, which is approximately 900 $\Omega^{-1}$cm$^{-1}$ for ZrMnP and 1400 $\Omega^{-1}$cm$^{-1}$ for HfMnP (**Figure 3(a)**)**.** It can be observed that the $\rho_{yz}^A$ vs. $\rho_{yy}^2$ plot down-turns slightly with an increasing temperature. The origin of such deviation is not very clear and it has also been observed in other compounds.[8, 12, 15, 19, 20] However, the shifting of the chemical potential with the temperature is the most typical and persuasive explanation, whereby the AHC is constant only through the SOC gap.[21]

.

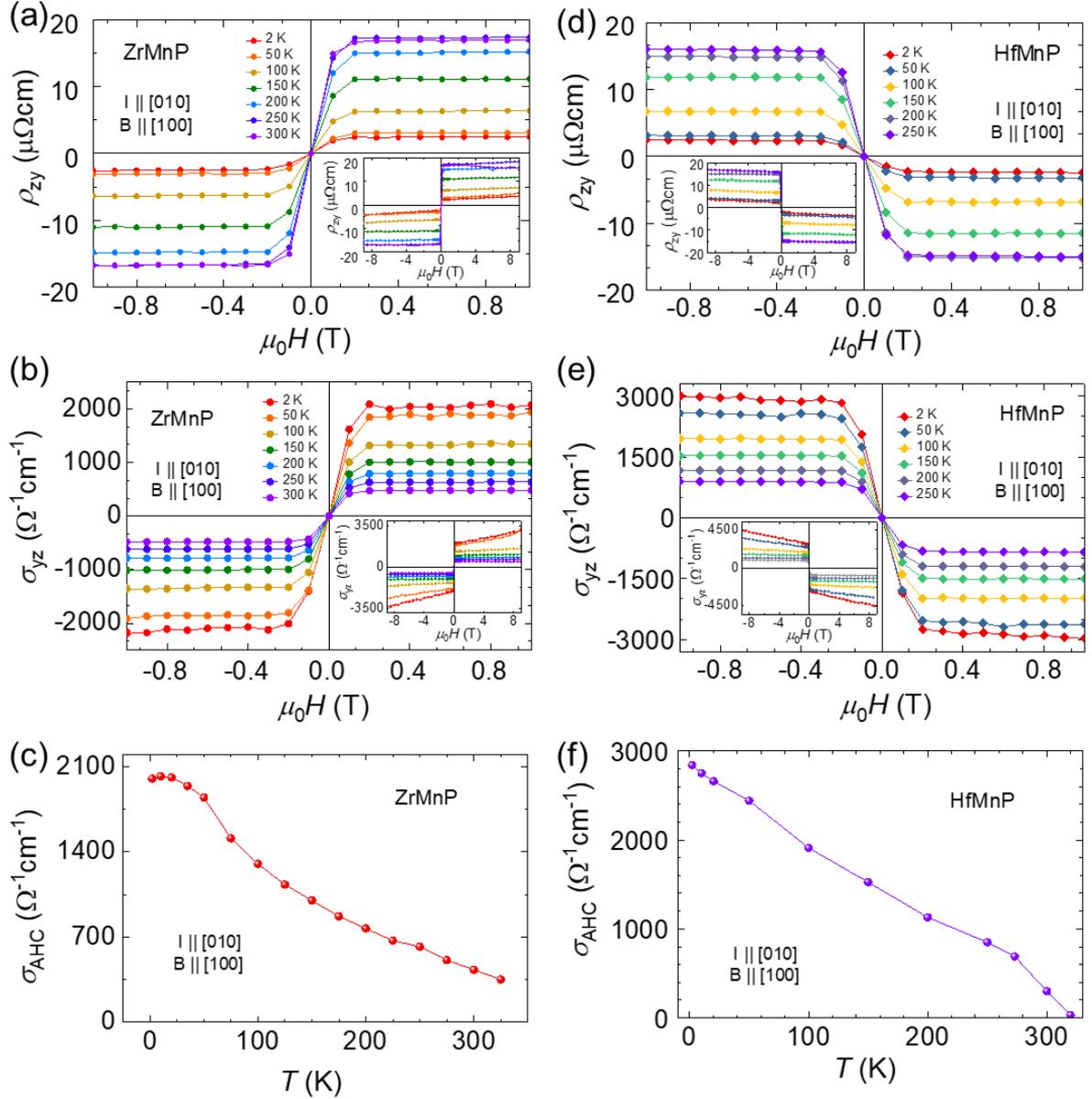

**Figure 2.** Hall resistivity $\rho_{yz}$ and conductivity $\sigma_{zy}$ of ZrMnP and HfMnP when field $B \parallel a$ and current $I \parallel b$ in the field range of ± 1 T. The left-column panels for ZrMnP: (a) field-dependent measured $\rho_{yz}$ at several temperatures and (b) corresponding calculated $\sigma_{zy}$ from relation $\sigma_{zy} = \frac{\rho_{yz}}{\rho_{yz}^2+\rho_{yy}^2}$, and (c) temperature-dependent extracted $\sigma_{AHC}$. The same measurements are provided in the right-column panels (d) to (f) for HfMnP. The corresponding insets are the data in the field range of ± 9 T.

In compounds that possess a large BC-induced AHC, a large anomalous Hall angle (AHA) is also expected. The AHA defines the ratio of the AHC to the longitudinal conductivity $\sigma_{yy}$ at the zero-field: AHA = $\sigma_{AHC}/\sigma_{yy}$ ($B$ = 0). The AHA exhibits a linearly increasing trend with an increase in the temperature, but decreases further after peaking at approximately 150 K. This is because as $T_C$ is approached, the anomalous behavior begins to diminish as the material starts to lose its ferromagnetic property. The ZrMnP and HfMnP exhibit very high AHA values of 10.2% and 13.6%, respectively (Figures 3(a) and (b)). These values are sufficiently high to maintain the present compounds among the various ferromagnetic nodal line compounds.[8, 12, 15, 19] Notably, the ZrMnP and HfMnP had the highest AHC values to the best of our knowledge as well as relatively high AHA values.

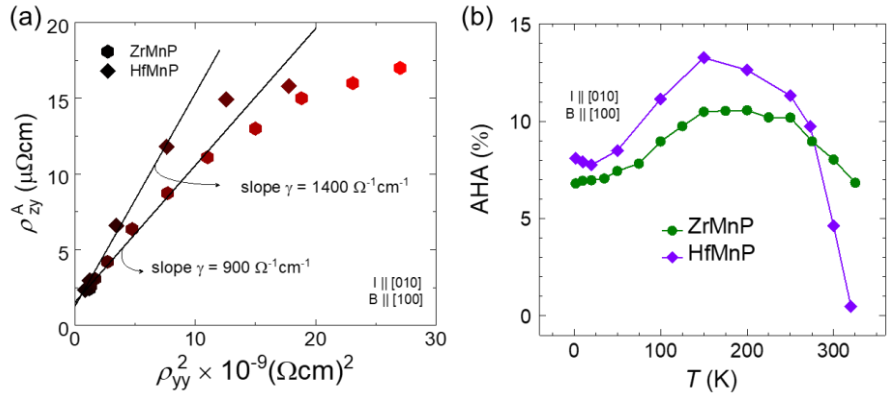

**Figure 3.** Unified scaling law to extract intrinsic AHC and AHA. (a) Plot of AHE vs. $\rho_{yy}^2$ at various temperatures, where the linear slope indicates the intrinsic AHC value. Refer to the text for the deviation from the linearity at the higher end. (b) Temperature-dependence measured AHA for ZrMnP and HfMnP, where AHA is defined as the ratio $\sigma_{AHC}/\sigma_{yy}$.

After measuring the AHC values, we applied the tight-binding method (see Methods section for details) to understand the origin of the AHC. The calculated band structures for the ZrMnP and HfMnP are presented in Figure 4(a) and (b), respectively. Except a tiny band along $\Gamma$-$Y$ (black line), only those bands (green lines) appear at the $E_F$, which contribute to the nodal-lines. Such bands primarily dominate in electrical transport. There are three mutually perpendicular mirror planes for the particular SG symmetry *Pnma*. Our calculations without SOC reveal several nodal loops that are located in the band structure, which are enforced by these mirror planes (Figure 4(c)). By considering the SOC and magnetization direction along the *a*-axis as in the experiments, the magnetic moments are not compatible with the mirror symmetry at $y = 0$.[8, 9] In this scenario, the nodal lines are no longer protected and the degeneracy is lifted (Figure 4(d)). This, in turn, creates strong BC contributions along the former nodal lines, as illustrated in Figure 4(e), which directly contributes to the intrinsic part of the AHC. The calculated intrinsic parts of the AHC are ~1000 $\Omega^{-1}$cm$^{-1}$ for ZrMnP and ~1500 $\Omega^{-1}$cm$^{-1}$ for HfMnP, which are in excellent agreement with the intrinsic parts of the experimental values. As these values are remarkably high, the studied compounds are interesting cases among the known nodal-line compounds.

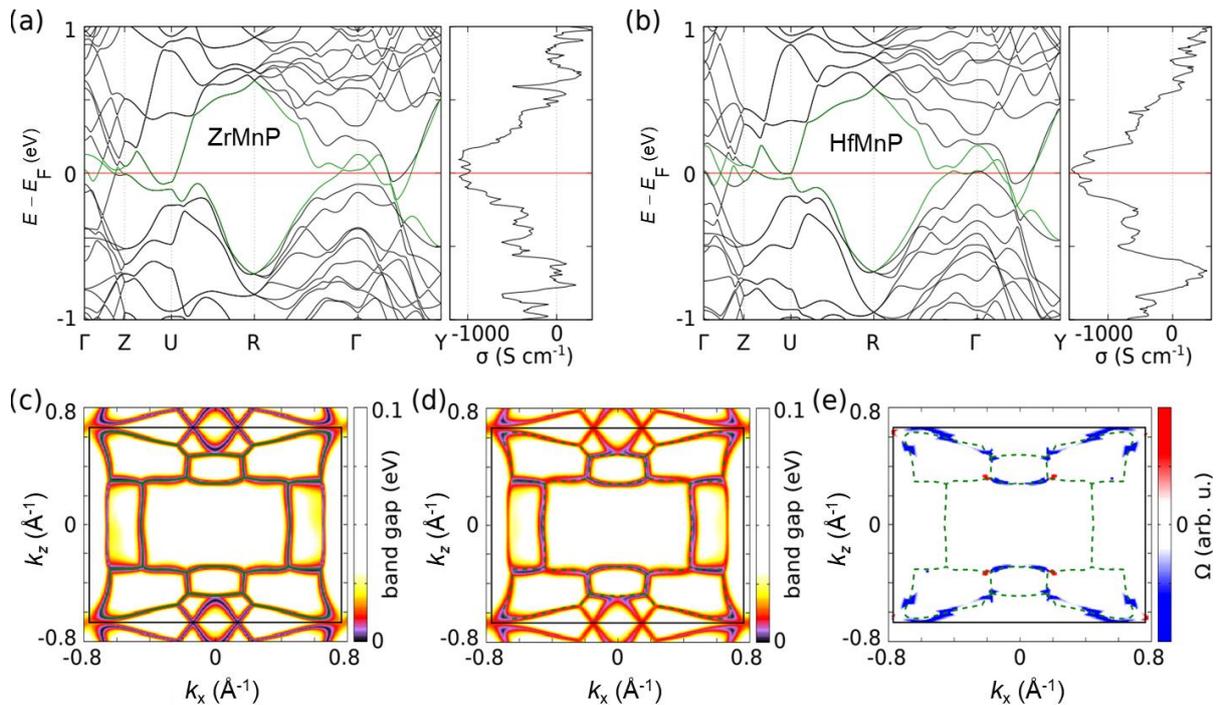

**Figure 4:** Band structure, AHC, and BC of ZrMnP and HfMnP. (a) Band structure and AHC of ZrMnP. (b) Band structure and AHC of HfMnP. In both (a) and (b), the bands (green lines) which contribute to forming nodal-lines (c) Band gap of bands marked in green of HfMnP in (b) in $y = 0$ plane without SOC. Several closed loops are visible (green lines). (d) $y = 0$ plane with SOC and magnetization along *a*-axis, with gapped nodal lines (dashed green lines). (e) BC, $\Omega$, at Fermi level in $y = 0$ plane. The large

contributions are located around the gapped nodal lines (dashed green lines). Similar results to those in (c) to (e) are also expected for ZrMnP.

Among the AHC values that were measured in different directions, the highest value was observed when $B \parallel a$. In this scenario, all nodal lines corresponding to the mirror plane at $y = 0$ has their degeneracy lifted and the nontrivial gap is opened, whereas the other nodal lines remain degenerate. As this gap is larger for HfMnP, we observed a larger AHC for HfMnP than for ZrMnP. Our previous study indicated that the number of mirror planes present in the compound plays a crucial role in the manipulation of the AHC.[10] This was demonstrated by using the simple example of two different space groups that possessed different numbers of mirror planes. SG 225 contains three mutually perpendicular mirror planes at $x = 0$, $y = 0$, and $z = 0$, whereas these planes are absent in SG 216. Consequently, the compounds that are associated with SG 225 exhibit larger AHC values compared to SG 216. Moreover, the location of the nodal line is an important factor. Naturally, the nodal line structure can only contribute to the transport if it is located at $E_F$. In these two cases, the number of mirror planes is related to the crystal structure, whereas the location is material specific. Our present investigation suggests that compound selection aimed at a high intrinsic value of AHC can be achieved by the tuning of the mirror symmetry that is inevitably present in various achiral space groups.

## 3. Conclusion

We have demonstrated the AHE in ferromagnetic ZrMnP and HfMnP. The crystal structure of these compounds comprises numerous mirror planes, which result in gapped nodal-line states in the band structure under the SOC effect. Among the various nodal lines, several lie at the Fermi energy, making these compounds unique compared to other known nodal-line compounds. These nodal lines are directly responsible for high AHC values owing to the accumulation of large BCs. Various advantages are offered by the present selection of compounds, such as a large ferromagnetic transition temperature, one of the largest ever intrinsic AHC values, a significantly large AHA, and the scope to observe the SOC effect on the AHC value. Our investigations provide various opportunities for tuning the AHC as well as searching for more compounds with higher numbers of mirror planes in their crystal structures.

## 4. Methods

*Crystal Growth:* The single crystals of ZrMnP and HfMnP were grown by the self-flux method, similar to that reported by Lamichhane et al.[17] Initially, Mn chips (99.99%) were cleaned by placement in an evacuated quartz tube, heated to 1000 °C at a rate of 200 °C/h, and maintained for 24 h before switching off the furnace. Shiny silver Mn pieces were obtained, which were subsequently ground into powder. For the single crystal growth, grounded Mn powder, P lumps (99.999%), and small pieces of Zr/Hf from their foils (99.8%) were used in a $(Zr/Hf)_{1.25}Mn_{85.9}P_{12.85}$ stoichiometry ratio and placed in the dried alumina crucibles. Thereafter, the crucibles were sealed in a quartz tube with 5 mbar partial pressure of argon. The entire reaction mixture was placed in a box-furnace and heated in two steps: first to 250 °C at a rate of 50 °C/h and maintained for 3 h, and subsequently to 1180 °C at the same heating rate. At 1180 °C, the reaction was maintained for 12 h for homogeneity. Slow cooling to 1025 °C was conducted over 180 h, at which time the additional flux was removed with centrifugation. Silver needle-shaped single crystals were obtained from the above temperature profile. The crystals of both the compounds were stable in air and moisture. Elemental analysis with EDX confirmed that the composition of the compounds was close to 1:1:1.

*Numerical Methods:* For the theoretical investigations, we employed *ab initio* calculations based on DFT, as implemented in VASP.[22] This code uses plane waves and pseudopotentials as a basis set. The exchange-correlation potential was used in the generalized gradient approximation.[23] The $k$ mesh used for the integration over the Brillouin zone was $7 \times 13 \times 7$. For calculations with a denser $k$ mesh, Wannier functions were extracted from the DFT results using the Wannier90 package.[24] Using these Wannier functions, we constructed a tight-binding Hamiltonian H, which was used to evaluate the BC $\Omega$ in the system as follows:[18, 21, 25]

$$\Omega = \sum_{m \neq n} \frac{\langle n|\frac{\partial H}{\partial k_i}|m\rangle\langle m|\frac{\partial H}{\partial k_j}|n\rangle - (i \leftrightarrow j)}{(E_n - E_m)^2},$$

where $|n\rangle E_n$ are the eigenstates and -energies of H. On this basis, the AHC was calculated as follows:[18, 21]

$$\sigma_{xy} = \frac{e^2}{\hbar} \sum_n \int \frac{d^3k}{(2\pi)^3} \Omega_{xy}^z f_n,$$

where $f_n$ is the Fermi distribution function. The $k$ mesh for the integration over the Brillouin zone in this step was selected as 301 × 301 × 301 to ensure converged results.

**Acknowledgements**

This work was financially supported by Deutsche Forschungsgemeinschaft (DFG) under SFB 1143 (Project No. 247310070), the European Research Council (ERC) Advanced Grant No. 742068 ("TOPMAT"), and Würzburg-Dresden Cluster of Excellence on Complexity and Topology in Quantum Matter—ct.qmat (EXC 2147, project no. 39085490).

**Conflict of Interest**

The authors declare no conflict of interest.

TOC:

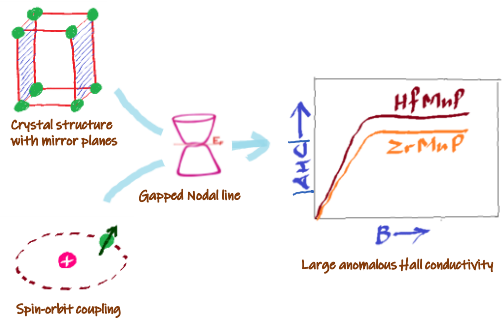

Anomalous Hall conductivity appears in metallic ferromagnets due to the anomalous velocity of charge carriers in the presence of magnetic field. In ferromagnets ZrMnP and HfMnP, the abundance of mirror planes forming nodal lines at the Fermi energy guides and enhances such velocity, resulting both the compounds exhibit large value of anomalous Hall conductivity.